\documentstyle[epsfig,preprint,aps]{revtex}
\begin{document}
\draft
\def\bra#1{{\langle #1{\left| \right.}}}
\def\ket#1{{{\left.\right|} #1\rangle}}
\def\bfgreek#1{ \mbox{\boldmath$#1$}}
\title{Electromagnetic Form Factors of the Nucleon in \break
an Improved Quark Model}
\author{D. H. Lu, A. W. Thomas, and  A. G. Williams}
\address{Department of Physics and Mathematical Physics\break
 and\break
        Special Research Centre for the Subatomic Structure of Matter,\break
        University of Adelaide, Australia 5005}
\maketitle
\vspace{-8.0cm}
\hfill ADP-97-16/T253
\vspace{8.0cm}
\begin{abstract}
Nucleon electromagnetic form factors are studied in the
cloudy bag model (CBM) with center-of-mass and  recoil corrections.
This is the first presentation of a full set of nucleon
form factors using the CBM.
The center of mass motion is eliminated via several different
momentum projection techniques and the results are compared. 
It is found that the shapes of these form factors are significantly 
improved with respect to the experimental data 
if the Lorentz contraction of the internal
structure of the baryon is also appropriately taken into account.
\end{abstract}

\section{Introduction}

Form factors characterise
the internal structure of subatomic particles and,  in particular,  
electromagnetic probes of hadrons  provide important
information on the underlying quark and gluon degrees of freedom.
In the nonperturbative regime (i.e., at low momentum transfer), 
QCD-motivated, effective hadronic  models continue to play an important role 
in analyzing and understanding a wealth of  experimental data.
The MIT bag model\cite{MIT} was an early attempt to include the key features 
of confinement and asymptotic freedom in a quark based model 
of hadronic structure.
The cloudy bag model (CBM)\cite{CBM} improves on the MIT bag model 
significantly by introducing an elementary pion field  coupled to the
quarks inside the bag  
such that chiral symmetry is restored.
The introduction of the pion field  not only improves the static 
nucleon properties, but also provides a  convenient connection to 
the study of conventional intermediate energy physics such as $\pi N$  and 
$NN$ scattering.  

There are many calculations of the nucleon electromagnetic form factors 
within different hadronic models.
Indeed the understanding of these form factors is extremely important in 
any effective theory or  model of the strong interaction.
However, there is, to our knowledge, no truly satisfactory means of 
forming fully
Lorentz covariant momentum eigenstates from any static model.
In this work we suggest an improved treatment (a hybrid method of Galilean
momentum projection combined with an appropriate
Lorentz contraction) for a  model which has been widely used for
many years  --- the CBM,
 and  bring it to larger momentum transfers.
The present results not only  remind us of the effectiveness 
of  chiral quark models at moderate momentum transfer, but also
remind us that they 
can serve as an essential first step in the investigation of the 
electromagnetic interaction in quark based nuclear models, 
in particular, the electromagnetic interaction 
in the quark-meson coupling (QMC) model\cite{medium97}.

In the CBM, as in the MIT bag model, 
quarks are independent particles confined in a rigid spherical well.
The bag model wave function for a baryon is a direct product of individual
quark wave functions, analogous to nuclear shell model wave functions.
A static bag cannot carry a definite momentum and so bag-model 
baryon states are not total momentum eigenstates, in spite of the fact 
that the Hamiltonian commutes with the total momentum operator.
Matrix elements evaluated between such states contain 
spurious center of mass motion.
This defect compromises some of the predictive power of the model, 
such that   
only observables involved in very low-momentum
transfer processes ($q^2/4m^2_N \ll 1$) are typically assumed to be reliable.

Over the years, a number of  prescriptions for the correction of 
the center of mass motion have been developed
(for an overview, see for example Refs.\cite{WILETS}). The diversity
of approaches may be viewed as an indication of  the uncertainty 
associated with this correction.
In contrast to the nonrelativistic case, the internal motion of a composite 
object cannot be explicitly separated from the collective motion 
in a covariant description. 
For the calculation of the form factors, a satisfactory treatment may result
from a combination of relativistic boost,
momentum projection, and a variational procedure.
Betz and Goldflam\cite{BETZ83} argued that a static soliton bag can be 
boosted consistently to a soliton bag moving with a finite velocity.
However this approach is impractical for boosting the  MIT bag because of 
the sharp surface
which prevents the construction of a simple boost operator\cite{ANTWERPEN94}.
A number of nonrelativistic methods for the center of mass correction
exist in the literature\cite{DJ80,WONG81,TEGEN82}. 
Analytic forms of the recoil corrections can be obtained  
in a relativitic harmonic oscillator quark model\cite{Tegen90}.
Unfortunately different groups do not always agree with each other 
and  sometimes 
even result in a correction with the opposite sign.  


In this work we compare 
several intuitively simple momentum-projection procedures
for the  calculation of the nucleon electromagnetic form factors. 
The basic idea is to extract the momentum eigenstates from the static solutions
by appropriate  linear superpositions. The simplest prescription for 
this approach was proposed by Peierls and Yoccoz (PY)\cite{PY57}.
We will assume that a baryon is composed of three constituents. Hence 
the wave function
for a moving baryon with total momentum ${\bf p}$ is constructed as
\begin{equation}
\Psi_{\rm{PY}}({\bf x}_1, {\bf x}_2, {\bf x}_3; {\bf p})
= N_{\rm{PY}}({\bf p})\int\! d^3{\bf x} e^{i {\bf p\cdot x}}
\Psi({\bf x}_1, {\bf x}_2, {\bf x}_3; {\bf x}), \label{PYWF}
\end{equation}
where $N_{\rm{PY}}({\bf p})$ is a momentum dependent normalization constant. 
The localized
state is simply given by a product of the three individual quark wavefunctions,
\begin{equation}
\Psi({\bf x}_1, {\bf x}_2, {\bf x}_3; {\bf x}) = 
q({\bf x}_1 - {\bf x}) q({\bf x}_2 - {\bf x}) q({\bf x}_3- {\bf x}),
\end{equation}
where  ${\bf x}$ refers to the location of the centre of the static bag, and 
${\bf x}_1$, ${\bf x}_2$, and ${\bf x}_3$ specify the positions 
of the three constituent quarks.
With the  PY wave function, the predictions of the static baryon properties
are generally improved\cite{WILETS}.
It reduces the r.m.s. radius, increases $g_A$, 
and on the whole produces a better mass spectrum. However, it is  unreliable 
for calculations of dynamic observables which involve
large momentum transfers, sine the PY wave function does not  transform 
appropriately under Lorentz boosts.

A closely related method for  eliminating the center of mass motion 
is called the  Peierls-Thouless (PT) projection\cite{PT62,DELTA}. There
the wave function is constructed through one further linear superposition
in terms of the PY wavefunction,
\begin{equation}
\Psi_{\rm{PT}}({\bf x}_1, {\bf x}_2, {\bf x}_3; {\bf p})
= N({\bf p}) \int\! d^3p' w({\bf p'}) e^{i({\bf p - p'}) \cdot {\bf x}_{\rm{c.m.}} }
\Psi_{\rm{PY}}({\bf x}_1, {\bf x}_2, {\bf x}_3; {\bf p'}).
\end{equation}
where ${\bf x}_{\rm{c.m.}} = ({\bf x}_1 + {\bf x}_2 + {\bf x}_3)/3$ 
is the center of mass of the baryon (we assume equal mass quarks here). 
Ideally the weight function, $w({\bf p'})$,  should be chosen to minimize
the total energy, but this is quite complicated to implement in practice.
As in Ref.~\cite{DELTA}, we make the choice $w({\bf p'}) = 1$ for simplicity
and convenience.  
Then integrations over  ${\bf x}$ and ${\bf p'}$ can be carried out 
explicitly.
This leads to a comparatively simple PT wave function for the baryon,
\begin{equation}
\Psi_{\rm{PT}}({\bf x}_1, {\bf x}_2, {\bf x}_3; {\bf p}) = 
N_{\rm{PT}} e^{i{\bf p \cdot x}_{\rm{c.m.}}}
q({\bf x}_1 - {\bf x}_{\rm{c.m.}}) q({\bf x}_2 - {\bf x}_{\rm{c.m.}})
q({\bf x}_3 - {\bf x}_{\rm{c.m.}}), \label{PTWF}
\end{equation}
where $N_{\rm{PT}}$ is determined  by the requirement that it satisfy 
the normalization  condition
\begin{equation}
\int\! d^3 x_1 d^3 x_2 d^3 x_3  
\Psi^\dagger_{\rm{PT}}({\bf x}_1, {\bf x}_2, {\bf x}_3; {\bf p'})
\Psi_{\rm{PT}}        ({\bf x}_1, {\bf x}_2, {\bf x}_3; {\bf p}) 
= (2\pi)^3 \delta^{(3)}(\bf p' - \bf p) \eta_p . \label{norm}
\end{equation}
with $\eta_p = 1$ for the nonrelativistic normalization and 
$\eta_p=E(p)/m_N$ if we wish to adopt a standard  
relativistic normalization for the baryon wavefunction.
%

Notice that the above methods of momentum projection 
act only  on the center of mass coordinate and the individual quark 
wavefunctions are not affected. 
However, since baryons are composite objects, once they
have nonzero momentum, their internal  structure should be subsequently 
modified.
For example, the bag surface is no longer spherical in the Breit frame,
rather it should be contracted along the direction of motion. 
We take care of this effect in terms of the prescprition by 
Licht and Pagnamenta\cite{LP70}.

It should be noted that the present work is the first 
presentation of calculations of the nucleon electromagnetic form
factors using the CBM, besides the obvious improvement of the treatment.
The outline of the paper is as follows. Firstly, we briefly review 
the electromagnetic interactions of the CBM in Sec.~II. The calculation of 
electromagnetic form factors for the bare bag with momentum projection is  
then presented in Sec.~III. 
In Sec.~IV, we discuss the necessary scaling of the form factors due to the 
effects of Lorentz contraction. Pionic corrections are then given in Sec.~V. 
The numerical results are presented and discussed  in Sec.~VI 
before the concluding remarks in Sec.~VII. 
Some technical details and explicit proof of gauge invariance 
of the calculations are provided in appendix.
 
\section{Electromagnetic currents in the CBM}
The linearized CBM Lagrangian with the pseudoscalar pion-quark coupling
(up to order $1/f_\pi$ ) is given by\cite{CBM} 
\begin{eqnarray}
        \protect{\cal L}
        &=&  (i\overline q \gamma^\mu \partial_\mu q - B)\theta_V
        - {1\over 2}\overline q q \delta_S \nonumber \\
        && + {1\over 2} (\partial_\mu \bfgreek{\pi})^2
        - {1\over 2} m^2_\pi \bfgreek{\pi}^2
        - {i\over 2f_\pi} \overline q \gamma_5 \bfgreek{\tau} \cdot
        \bfgreek{\pi} q \delta_S, \label{LAG}
\end{eqnarray}
where $B$ is a bag constant, $f_\pi$ is the $\pi$ decay constant, 
$\theta_V$ is a step function (unity inside the bag volume
and vanishing  outside) and
$\delta_S$ is  a surface delta function.
In a lowest order perturbative treatment of the pion field,
 the quark wave function is not affected by the pion field and 
is simply given by the MIT bag solution\cite{MIT}
\begin{equation}
q({\bf r}) =  \left (
\begin{array}{c} g(r) \\ i\bfgreek {\sigma} \cdot\hat{r} f(r) \end{array} \right
 ) \phi\, \theta(R-r),
\end{equation}
where $\phi$ contains the spin-isospin information for the wavefunction 
of the quark, 
${\bfgreek {\sigma}}$ is the usual  Pauli spin operator, and 
$R$ is the spherical bag radius.
For the ground state of a massless quark
$g(r) = N_s j_0(\omega_s r/R), f(r) = N_s j_1(\omega_s r/R) $,
where  $\omega_s = 2.0428$ and
 $N_s^2 = \omega_s/8\pi R^3 j_0^2(\omega_s) (\omega_s -1)$.

From the CBM Lagrangian given in  Eq.\ (\ref{LAG}), the conserved local  
electromagnetic current can be derived using 
 the principle of minimal coupling 
$\partial_\mu \rightarrow \partial_\mu + i q A_\mu$,
where $q$ is the charge carried by the field upon
which the derivative operator acts. 
The total electromagnetic current is then
\begin{eqnarray}
J^\mu(x) &=& j^{\mu(Q)}(x) + j^{\mu(\pi)}(x), \label{current} \\
j^{\mu(Q)}(x) &=& \sum_f Q_f e \overline{q}_f(x) \gamma^\mu q_f(x), \\
j^{\mu(\pi)}(x) &=& -i e [ \pi^\dagger(x) \partial^\mu \pi(x)
               -\pi(x) \partial^\mu \pi^\dagger(x)],
\end{eqnarray}
where $q_f(x)$ is the quark field operator 
for the flavor $f$, $Q_f$ is its charge in units of $e$, and $e \equiv |e|$ 
is the magnitude of the electron charge.
The charged pion field operator is defined as
\begin{equation} 
 \pi(x) = {1\over \sqrt{2}}[\pi_1(x) + i\pi_2(x)],
\end{equation}
which either destroys a negatively charged pion
or creates a positively charged one.

The physical baryon
state is then a  dressed bag, consisting of a superposition of 
a bare bag and a bag with a pion cloud. Algebraically, it has the  form
\begin{equation}
\ket A = \sqrt{Z_2^A} [ 1 + (m_A - H_0 - \Lambda H_I \Lambda )^{-1} H_I ] 
\ket {A_0} \label{state},
\end{equation}
where $Z^A_2$ is the bare baryon probability in the physical baryon states,
\begin{equation}
Z^A_2 = \left[ 1 +
 \sum_B \left({f^{AB}_0\over m_\pi}\right)^2 {1\over 12\pi^2}
\, \mbox{P}\!\int_0^\infty\! {dk\, k^4 u^2(k R)\over 
\omega_k (m_A - m_B - \omega_k)^2}
\right]^{-1},
\end{equation}
where $\Lambda$ is a projection operator which 
annihilates all the components of $\ket A $ without at least one pion, 
and $H_I$ is the interaction Hamiltonian
which describes the process of emission and absorption of pions. 
We follow the traditional CBM treatments and consider only states with at most
one pion. The matrix elements of $H_I$ between the bare baryon states and 
their properties are then given by\cite{TT83}
\begin{eqnarray}
v^{AB}_{0j}(\vec{k}) &\equiv& 
\bra {A_0} H_I \ket{{\bf \pi}_j(\vec{k}) B_0} = {i f^{AB}_0\over m_\pi}
{u(kR) \over [2\omega_k (2\pi)^3]^{1/2}} \sum_{m,n}
C^{s_B m s_A}_{S_B 1 S_A} (\hat{s}^*_m \cdot {\vec k}) 
C^{t_B n t_A}_{T_B 1 T_A} (\hat{t}^*_n \cdot {\vec e}_j),\\
w^{AB}_{0j}(\vec{k}) &\equiv& 
\bra{A_0 {\bf \pi}_j(\vec{k})} H_I \ket {B_0}
 = \left[v^{BA}_{0j}(\vec{k})\right]^* 
= -v^{AB}_{0j}(\vec{k}) = v^{AB}_{0j}(-\vec{k}),
\end{eqnarray}
where the pion has momentum $\vec{k}$ and  isospin projection $j$.
Note also  that 
 $f_0^{AB}$ is the reduced matrix element for the 
$\pi B_0 \rightarrow A_0$ transition vertex, $u(kR) \equiv 3j_1(kR)/kR $,
$\omega_k = \sqrt{k^2+ m^2_\pi}$, and $\hat{s}_m$ and $\hat{t}_n$ 
are spherical unit vectors for spin and isospin, respectively.

\section{Momentum Projection Calculations for A Bare Bag}
It is customary to define the nucleon electric ($G_E$) and magnetic ($G_M$)
form factors in the Breit frame by
\begin{eqnarray}
\bra {N_{s'}({\vec{q}\over 2}) }  J^{0}(0) \ket {N_s(-{\vec{q}\over 2})} 
&=& \chi_{s'}^\dagger\,\chi_s\, G_E(q^2), \label{GE} \\
\bra {N_{s'}({\vec{q}\over 2}) } \vec{J}(0) \ket {N_s(-{\vec{q}\over 2})} &=& 
\chi_{s'}^\dagger\, {i\bfgreek\sigma\times\vec{q}\over 2 m_N}\,\chi_s\, 
G_M(q^2), \label{GM}
\end{eqnarray}
where $\chi_s$ and $\chi^\dagger_{s'}$ are Pauli spinors for 
the initial and final nucleons, $\vec{q}$ is the Breit-frame 
three momentum transfer, 
i.e., $q^2 = q_0^2 - \vec{q^2} = -\vec{q^2} = -Q^2$. 
We choose $\vec{q}$ to define the z-axis. 
The major advantage of the Breit frame is that $G_E$ and $G_M$
are explicitly  decoupled, and can be determined respectively 
by the time and space
components of the electromagnetic current operator $J^{\mu}$.

In the definition above [i.e., Eqs.~(\ref{GE}) and (\ref{GM})], 
both initial and final states 
are physical states. Using Eqs.~(\ref{current}) and (\ref{state}), the total
electromagnetic form factors can be expressed  in terms of the three processes 
 shown in Fig.~1.
In this section we calculate the contribution 
from the bare bag only, and leave 
the pion loop effects to be  included in a later section. 
For the three-momentum eigenstates, Eqs.~(\ref{PYWF}) and (\ref{PTWF}), 
we can proceed to calculate the electromagnetic form factors 
in a relatively straightforward way. 
For the PY projection, we obtain,
\begin{eqnarray}
G^{\rm PY}_E(q^2) &=& I_E(q^2) / D_{\rm PY}(q^2), \label{PYE}\\
G^{\rm PY}_M(q^2) &=& I_M(q^2) / D_{\rm PY}(q^2), \label{PYM}
\end{eqnarray}
where 
\begin{eqnarray}
I_E(q^2) &=& \int_0^\infty\!\! dz z^2 \, N_Q^2(z) \int\!\! d^3r\, j_0(q r)
\left[ g_+g_- + {f_+f_- \over r_+r_-} \left(r^2 - {z^2\over 4}\right)\right]
\Theta,\\
I_M(q^2) &=& 2 m_N \int_0^\infty\!\! dz z^2 \, N_Q^2(z) \int\!\! d^3r\, 
{j_1(q r)\over q r}\\ \nonumber
&&
\left [ r^2 \left ( {g_-f_+ \over r_+} + {g_+f_- \over r_-} \right ) 
+ {\vec{r}\cdot\vec{z} \over 2} 
\left ( {g_-f_+ \over r_+} - {g_+f_- \over r_-}\right ) \right ]
\Theta,\\
D_{\rm PY}(q^2) &=& \int_0^\infty\! dz z^2 \, N_Q^3(z)\, j_0(qz/2)\Theta,\\
N_Q(z) &=& \int\! d^3r\, \left [ g_+g_- + {f_+f_-\over r_+r_-}
\left ( r^2 - {z^2\over 4}\right ) \right ] \Theta.
\end{eqnarray}
Here $D_{\rm PY}(q^2)$ is the momentum dependent normalization factor 
and $N_Q(z)$ is the
overlap integral associated with each quark spectator.
The following shorthand notation has been used in the above equations,
\begin{eqnarray}
r_\pm      &\equiv& \left| \vec{r} \pm {\vec{z} \over 2}\right| ,\\
g_\pm      &\equiv& g(r_\pm), \,\,\, f_\pm \equiv f(r_\pm),\\
\Theta &\equiv& \theta(R - r_+)\theta(R - r_-).
\end{eqnarray}

Similarly, for the PT projection, we obtain
\begin{eqnarray}
G_E^{\rm{PT}}(q^2) &=& \int\! d^3r j_0(qr)\rho(r)K(r)/D_{\rm PT}, \label{PTE}\\
G_M^{\rm{PT}}(q^2) &=& \int\! d^3r j_1(qr)g(r) f(r)K(r)/D_{\rm PT}, \label{PTM}
\end{eqnarray}
where
\begin{equation}
D_{\rm PT} = \int\! d^3r \rho(r) K(r),
\end{equation}
with $\rho(r) \equiv g^2(r) + f^2(r)$ and 
$K(r) \equiv \int d^3x \rho(\vec{x}) \rho(-\vec{x} - \vec{r})$
is the recoil function to account for the correlation of the 
two spectator quarks.

As expected, without the momentum projection, 
 Eqs.~(\ref{PYE}, \ref{PYM}) and Eqs.~(\ref{PTE}, \ref{PTM}) would reduce to
the familiar results  for the static, spherical MIT bag, i.e.,  
\begin{eqnarray}
G_E^{(\rm static)}(q^2) &=& \int d^3r\, j_0(q r)\, [g^2(r) + f^2(r)], \\
G_M^{(\rm static)}(q^2) &=& 2 m_N \int d^3r\, {j_1(q r)\over q }\,  
[2 g(r) f(r)] .
\end{eqnarray}
Note that nonrelativistic normalization of the nucleon wave functions has been
used here [see Eq.~(\ref{norm})], 
as is appropriate for the simple (Galilean) three-momentum projections
being used here.

\section{Corrections from the Lorentz contraction}

A complete solution of a covariant
many-body problem is extremely difficult. There is a substantial body 
of literature which uses light-cone dynamics\cite{LC} 
for the constituent quarks. With a few parameters this approach
can reproduce experimental data over quite a large momentum transfer range.
For the bag model, there is no Lorentz covariant solution for 
an extended quantum object in 
more than two dimensions\cite{Jaffe81}.  
Thus  we use a semiclassical prescription here.

As mentioned in the introduction, the spherical bag 
is expected to undergo a Lorentz contraction along the direction of motion
once it acquires a momentum.
An intuitive prescription by Licht and Pagnamenta\cite{LP70} 
suggested that, in the preferred Breit frame, 
the interaction of the individual
constituents of a cluster with the projectile may be regarded as instantaneous
to a good approximation. Relativistic form factors can be simply
derived from the corresponding nonrelativistic ones 
by a simple substitution rule.
In the case of the bag model, once the spurious center of mass motion 
is subtracted using the PY or PT procedure, 
it is natural to rescale the quark internal coordinates as well, i.e., 
\begin{equation}
\Psi(\vec{x}_1,\vec{x}_2,\vec{x}_3; \vec{0}) 
\stackrel{\rm projection}{\longrightarrow} 
\Psi(\vec{x}_1,\vec{x}_2,\vec{x}_3; \vec{p}) 
\stackrel{\rm contraction}{\longrightarrow}
\Psi(\vec{x'}_1,\vec{x'}_2,\vec{x'}_3; \vec{p})
\end{equation}
where the quark coordinates $\vec{x}'_i$ for the moving bag
 are related to the $\vec{x}_i$ 
by a Lorentz transformation. Without loss of generality, we again choose the
photon momentum $\vec{q}$ along the $z$-direction. Then in the Breit frame,
the quark displacements must contract in the $z$-direction 
while they remain unchanged in the $x$ and $y$ directions. Thus we have
\begin{eqnarray}
z'_i &=& {m_N\over E} z_i, \\
d^3x'_1 d^3x'_2 &=& ({m_N\over E})^2 d^3x_1 d^3x_2,
\end{eqnarray}
where we have assumed $t=0$ for all constituents, i.e., 
the instantaneous approximation. Note that $m_N$ 
is the nucleon mass and $E$ is the on-shell nucleon energy in the Breit frame.
The   $(m_N/E)^2$ factor is due to the Lorentz contraction of the coordinates
of the two spectator
quarks along the direction of motion.
As an example, the proton charge form factor  in the PT scheme 
is thus
\begin{eqnarray}
G_E(q^2) &=& \int\! \mathop{\prod}_i d^3x'_i e^{-iqz'_3}
[q_{\vec{p}}^\dagger(\vec{x'}_3-\vec{x'}_{\rm c.m.})
\delta(\vec{x'}_3) q_{\vec{p}}(\vec{x'}_3-\vec{x'}_{\rm c.m.})]\, 
\rho_{\vec{p}}(\vec{x'}_1-\vec{x'}_{\rm c.m.})
\rho_{\vec{p}}(\vec{x'}_2-\vec{x'}_{\rm c.m.}) \nonumber\\
&=& ({m_N\over E})^2\int\! d^3x_1 d^3x_2 e^{-iqz (m_N/E)}
\rho(\vec{x}_1-\vec{x}_{\rm c.m.})
\rho(\vec{x}_2-\vec{x}_{\rm c.m.})
\rho(-\vec{x}_{\rm c.m.})\nonumber\\
&=& ({m_N\over E})^2 G^{\rm sph}_E(q^2 m_N^2/E^2), \label{formula}
\end{eqnarray}
where $q_{\vec{p}}$ is the quark wave function in the Breit frame 
(in a deformed bag), 
$\rho_{\vec{p}}$ is the probability density of the quark, 
and $G^{\rm sph}_E$ is the charge form factor calculated 
with the spherical static bag wave function 
[such as Eqs.(\ref{PYE}) and (\ref{PTE})].
In the second step of the derivation we have used the fact that 
a probability amplitude is a constant in different Lorentz frames, 
hence, the identity $q_{\vec{p}}(\vec{x'}) \equiv q(\vec{x})$ 
has been used as in Ref.\cite{LP70}.
For the magnetic form factor, a similar expression can be derived,
$G_M(q^2) = ({m_N/E})^2 G^{\rm sph}_E(q^2 m_N^2/E^2)$.
Note that we used the fact that all three quarks have the same 
spatial wave function in obtaining Eq.~(\ref{formula}).
The scaling factor in the argument is due to the coordinate change of
the struck quark and
the factor in the front, $(m_N/E)^2$,  comes from  the reduction 
of the integral measure of two spectator quarks in the Breit frame.  
Note that this prescription is similar, but not identical, to the Lorentz
contraction arguments used in the Skyrme model\cite{JI91}. 
The difference is the $(m_N/E)^2$ factor in front of $G_E^{\rm sph}$ which is
absent in Ref.~\cite{JI91}.
It might be argued that, since we use only a nonrelativistic momentum
projection [i.e., $\eta_p$ in Eq.~(\ref{norm}) cannot be fixed unambiguously],
this factor is not well determined. 

\section{Corrections from the perturbative pion cloud}

In the CBM with a bag radius above 0.7 fm 
the pion field is  relatively weak and  the pionic effects
can be included perturbatively\cite{DODD81}.  As usual we assume that there is 
{\em no more than one pion in the air}.
There are two processes contributing to the nucleon 
electromagnetic form factors due to the pion cloud. 
One involves the direct $\gamma\pi\pi$ 
coupling shown in Fig.~1(c), and the other is the $\gamma q q$ coupling
inside a pion loop, as in Fig.~1(b).

Fig.~1(c) actually contains  three time-ordered subdiagrams, 
and has been evaluated in the CBM by Th\'eberge and Thomas\cite{TT83}. 
They gave 
\begin{equation}
G_{(E,M)}^{(\pi)}(q^2) = G_{(E,M)}^{(\pi)}(q^2; N) +  
	                 G_{(E,M)}^{(\pi)}(q^2; \Delta)
\end{equation}
where the two terms  correspond to two cases with 
different intermediate baryons ($N$ and $\Delta$). For completeness,
we quote the individual contributions explicitly,  
\begin{eqnarray}
G_{E}^{(\pi)}(q^2; N) &=& {1\over 36\pi^3}\!
\left ({f^{NN}\over m_\pi}\right )^2 
\!\!\int\! d^3k\, {u(kR)\, u(k' R)\, \vec{k}\cdot\vec{k'}\over
\omega_k \omega_{k'} (\omega_k + \omega_{k'}) }\, \bra N\tau_3\ket N ,
\label{EN}\\
G_{E}^{(\pi)}(q^2; \Delta) &=& 
-{1\over 72\pi^3}\!\left ( {f^{N\Delta}\over m_\pi}\right )^2
\!\!\int\! d^3k\, {u(kR)\, u(k' R)\, \vec{k}\cdot \vec{k'}\over
(\omega_{\Delta N} + \omega_k) (\omega_{\Delta N} + \omega_{k'}) 
(\omega_k + \omega_{k'})}\, \bra N \tau_3 \ket N , 
\label{ED}\\
G_{M}^{(\pi)}(q^2; N) &=& {2 m_N\over 72\pi^3}\! \left ({f^{NN}\over m_\pi}\right )^2 
\!\!\int\!\! d^3k\, {u(kR)\, u(k' R)\, (\hat{q}\times\vec{k})^2\over
(\omega_k \omega_{k'})^2}\, \bra N \tau_3 \ket N ,
\label{MN}\\
G_{M}^{(\pi)}(q^2;\Delta) &=& 
{2 m_N\over 288\pi^3}\!\left ( {f^{N\Delta}\over m_\pi}\right )^2
\!\!\int\!\! d^3k\, {(\omega_{\Delta N} + \omega_k + \omega_{k'})\, u(kR)\, u(k' R)\, 
(\hat{q}\times\vec{k})^2\over
(\omega_{\Delta N} + \omega_k) (\omega_{\Delta N} + \omega_{k'}) 
(\omega_k \omega_{k'})(\omega_k + \omega_{k'})}\, \bra N \tau_3 \ket N ,
\label{MD}
\end{eqnarray}
where $\vec{k}' = \vec{k} + \vec{q}$, $\omega_{BN}\simeq m_B - m_N$, 
$f^{NB}$ is the renormalized $\pi NB$ coupling constant,
and $\tau_3$ is the third nucleon isospin Pauli matrix.

Corresponding to Fig.~1(b), the transition matrix element can be written as
\begin{eqnarray}
\langle N({\vec{q}/2})\left|\right. &j^{\mu(Q)}(0)&
\left.\right| N(-{\vec{q}/2})\rangle =
\sum_{BC,j}\int\! d^3k \nonumber\\ 
& \times &
{ \bra{N({\vec{q}\over 2})} H_I\ket{N(\vec{p'}),\pi_j(\vec{k})}
\bra{B(\vec{p'})} j^{\mu(Q)}(0) \ket{C(\vec{p})}
\bra{C(\vec{p}),\pi_j(\vec{k})} H_I \ket{N(-{\vec{q}\over 2})} \over 
(\omega_{BN} + \omega_k)(\omega_{CN} + \omega_k) }, \label{f1b}
\end{eqnarray}
where $\vec{p'} = (\vec{q}/ 2) + \vec{k}$ and 
$\vec{p} = -(\vec{q}/ 2) + \vec{k}$ are the momenta for the intermediate
baryons $B$ and $C$.
With the dynamical baryons and pion here, we have to evaluate the 
electromagnetic  matrix elements for the intermediate processes
in an arbitrary frame. Thus the matrix elements of $J^{0}$ might contain
 both $G_E(q^2)$ and $G_M(q^2)$, as do those of  $\vec{J}$.
It is convenient to use the identity
\begin{equation}
\bra{p'}j^{\mu(Q)}(0)\ket{p} = \bar{u}(p')\left[\gamma^\mu F_1(q^2) + 
{i\sigma^{\mu\nu}q_\nu \over 2m_N}F_2(q^2)\right] u(p),
\label{dirac}
\end{equation}
where $F_1(q^2) = (G_E(q^2) + \eta G_M(q^2))/(1+\eta)$ and 
      $F_2(q^2) = (G_M(q^2) - G_E(q^2))/(1+\eta)$ with $\eta = -q^2/4m_N^2$.
Both $F_1(q^2)$ and $F_2(q^2)$  are Lorentz scalar functions
and hence can be evaluated in any frame. 
However, it can be shown that after integrating over the loop momentum, 
$\vec{k}$, the time ($G_E$) and space ($G_M$) components of Eq.~(\ref{f1b})
decouple again as long as the overall matrix element is evaluated  
in the Breit frame. The detailed expressions are messy and  
are therefore given in the appendix.

\section{Results and Discussion}
In this work, we have adopted the usual philosophy for the renormalization
in the CBM, using the approximate relation,
$ f^{AB} \simeq \left({f^{AB}_0\over f^{NN}_0}\right) f^{NN}$.
There are uncertain corrections on the bare coupling constant $f^{NN}_0$,
such as the nonzero quark mass and the correction for spurious 
center of mass motion.
Therefore, we use the renormalized coupling constant in our
calculation, $f^{NN} \simeq 3.03$, 
which corresponds to the usual $\pi NN$ coupling constant, 
$f^2_{\pi NN}\simeq 0.081$.



It should  point out that there is no unambigious way to implement 
strict gauge invariance (the Ward-Takahaski identities) for
a composite particle\cite{Gross}. In this work, we ensure a somewhat weak 
requirement -- electromagnetic current conservation\cite{Miller97}.
The explicit proof is given in the appendix.
Recall that Eqs.~(\ref{EN}), (\ref{ED}), (\ref{MN}), and (\ref{MD}) 
are results evaluated under a heavy baryon approximation. 
Ideally, Fig.~1(c) should be evaluated on the same footing as Fig.~1(b).
Numerical calculations show that the recoil effects for intermediate baryon
in the pion loop are negligible, and  we may therefore ignore this recoil 
and use the standard static CBM results for the pionic correction.
Consequently, the charge form factors at zero momentum transfer
automatically satisfy the
requirement of charge conservation, i.e.,
 $G_E(0) = G_{E}^{(Q)}(0) + G_{E}^{(\pi)}(0) = e_N$, 
where $e_N$ is 1 for the proton and 0 for the neutron.

The magnetic moments are simply the values of the magnetic form factors
at zero momentum transfer, $\mu \equiv G_M^{(Q)}(0) + G_M^{(\pi)}(0)$.
Note that the expression for the contribution from Fig.~1(b)
is somewhat different from Ref.~\cite{TT83}.
Here, there is no $Z_2$ factor for Fig.~1(b) in consistence with 
the charge conservation.
As a result of this choice the numerical contribution from  Fig.~1(b) 
increases by roughly $30\%$ 
($Z_2^N \simeq 0.73$ for R = 1 fm ), 
bringing  the  total nucleon magnetic moments a few percent closer to the 
experimental data. 

Table I gives the nucleon magnetic moments 
in this calculation. The center of mass correction reduces the static values 
by $5-10 \%$. This is in contradiction with Ref.~\cite{DJ80} but is
consistent with Refs.~\cite{TEGEN82,KAZUO89}.
The bag radius dependence is significantly reduced by the pion cloud 
in the CBM, with little variation over the range $R = 0.8 - 1.0$ fm. 
The deficiency of the nucleon magnetic moments may be attributed to
the higher order pionic corrections and explicit vector meson contributions.

The difference between the two choices of normalization of the wave function 
[i.e., the factor $\eta_p = 1$ or $E/m_N$ in Eq.~(\ref{norm}) ] 
is not significant with respect to
the shape of the form factors. However it will smoothly 
scale these form factors.
 For the proton charge form factor, for example,
the relativistic normalization raises the form factor roughly $5\%$ 
at $Q^2 = 0.5 \mbox{ GeV}^2$ and $10\%$ at $Q^2 = 1.0 \mbox{ GeV}^2$. 
For clarity, we have always used $\eta_p = 1$ in the following figures
as previously stated.

The characteristic effect of the center of mass correction on the 
charge form factor of the bare proton bag is illustrated 
 in Fig.~\ref{fig2.ps} with the bag radius R = 1 fm.
The ``dipole'' refers to the standard dipole fit, 
$F(Q^2) = 1/(1 + Q^2/0.71 \mbox{ GeV}^2)^2$.
The  bare charge form factors calculated with the static bag 
usually drop too quickly.   
With the PY projection procedure, the form factor at moderate momentum 
transfer ($Q^2 \sim 0.5 \mbox{ GeV}^2$) increases nearly $100\%$. 
However the shape of the form factor
does not change very much. 
It is generally too stiff and drops too fast which is mainly
due to the sharp surface of the cavity approximation
and lack of translational invariance of the wave function.
Using  the translational invariant PT projection procedure 
leads to  improved behavior of the form factors. 
In particular, after including the correction arising from 
Lorentz contraction, the shape of the form factors
is significantly improved.  It is reassuring to see that the combination of
Lorentz contraction and Galilean (nonrelativistic) momentum projection
is less scheme dependent than the momentum projection alone, ---e.g.,
compare the pairs of curves PY and PT with PYL and PTL in Fig.~2.  

Fig.~\ref{fig3.ps} and Fig.~\ref{fig4.ps} 
show the individual  contributions 
to the charge form factors from the quarks and the pion cloud,
for a typical bag radius of R = 1 fm. 
For the proton charge form factor,
it is clear that the $\gamma qq$ coupling terms dominate
the form factor where the bare photon--bag coupling contributes nearly
$75 \%$. The correction from the $\gamma\pi\pi$ coupling decreases 
very quickly as the momentum transfer increases. 
A smaller bag radius will lead to  a larger pionic contribution.
For the neutron charge form factors, the contribution from 
the photon--bare bag coupling [Fig.~1(a)]
vanishes due to the SU(6) structure. The severe cancellation between 
Fig.~1(b) and Fig.~1(c) results in a small but nonvanishing 
neutron charge form factor, with a negative mean square radius 
as one would expected simply
from the Heisenberg Uncertainty Principle. 
With the bag radius $R=1$ fm, we obtain the neutron charge r.m.s. radius 
of $\left<r^2\right>_{En} = -0.14 \mbox{ fm}^2$, to be compared with
the experimental value of $-0.12 \mbox{ fm}^2$\cite{nrms}. 

Figs.~\ref{fig5.ps}--\ref{fig8.ps} show the bag radius
dependence of the nucleon electromagnetic form factors. A large bag radius
always leads to a  softer form factor. We have used PT wave functions
with Lorentz contraction in these calculations.
The predictions show quite a reasonable agreement with the experimental data
to much larger values of the momentum transfer than one 
has tended to expect.
%

\section{summary}
We have calculated the electromagnetic form factors of the nucleon 
within the CBM, including relativistic corrections in the form of  
momentum projection and the Lorentz contraction of the internal structure.
Electromagnetic current conservation is ensured in this
calculation which is performed in the Breit frame.
This is the first time that a presentation of all the nucleon
electromagnetic form factors has been made for the CBM.
The two different procedures of momentum projection for the spurious 
center-of-mass motion give results which are relatively close to each other
when Lorentz contraction effects are included.
The Galilean invariant PT projection is  generally 
 a little better than the PY method in that it leads to a shape more 
closely resembling the dipole form.
Including  the corrections for  center of mass motion and 
 Lorentz contraction, the numerical predictions are in rather good agreement 
with data in the region $Q^2 < 1 \mbox{ GeV}^2$. 
This is quite a remarkable result when one realizes the simplicity of the
model. In particular, there are no explicit vector meson contributions
and one possible future development would be to include 
$\pi\pi$ interactions.

The authors are very grateful to I.R. Afnan and K. Tsushima
for valuable discussions. 
We also thank S. Platchkov for sending the data for 
the neutron charge form factor.
This work was supported by the Australian Research Council.


\begin{appendix}
\section{}

In Fig.~1(b), the intermediate $\gamma BC$ vertex can no longer 
be in the Breit frame. A strightforward evaluation gives the matrix elements
\begin{eqnarray}
\bar{u}(p')\gamma^0 u(p) &=& 
N'N\left[ 1+{k^2-q^2/4\over (E'+m_N)(E+m_N)}\right] + {\cal O}(\vec{k}), \\
\bar{u}(p'){i\sigma^{0\nu}q_\nu \over 2m_N} u(p) &=&
-{N'N\over 2m_N}\left( {q^2/2+\vec{k}\cdot\vec{q}\over E'+m_N}      
 -{-q^2/2+\vec{k}\cdot\vec{q}\over E+m_N}\right) + {\cal O}(\vec{k}), \\
\bar{u}(p')\gamma^a u(p) &=& 
{i\over 2}(\bfgreek{\sigma}\times\vec{q})_a N'N
\left({1\over E'+m_N}+{1\over E+m_N}\right) + {\cal O}(\vec{k}), 
\label{space1} \\
\bar{u}(p'){i\sigma^{a\nu}q_\nu \over 2m_N} u(p) &=&
{i\over 2m_N}(\bfgreek{\sigma}\times\vec{q})_a N'N
\left[1-{q^2/4+k^2\over (E'+m_N)(E+m_N)} \right] + {\cal O}(\vec{k}), 
\label{space2}
\end{eqnarray}
where $\vec{p'} = (\vec{q}/ 2) + \vec{k}$, 
$\vec{p} = -(\vec{q}/ 2) + \vec{k}$,
$E=(\vec{p}^2+m_N^2)^{1/2}$, $E'=(\vec{p'}^2+m_N^2)^{1/2}$, 
and the normalization constants are  $N=[(E+m_N)/2m_N]^{1/2}$, 
$N'=[(E'+m_N)/2m_N]^{1/2}$. 
The index $a$ in Eqs.~(\ref{space1}) and (\ref{space2}) 
denotes a space component, the ${\cal O}(\vec{k})$ terms in all equations 
refer to other pieces which are odd in $\vec{k}$ and 
will vanish after integration over the loop momentum $\vec{k}$. 
Substituting the above matrix elements into Eq.~(\ref{f1b}) 
and performing some spin and isospin algebra, we obtain
the nucleon electric and magnetic form factors originating 
from the $\gamma qq$ coupling
[i.e., the combination of  Fig.~1(a) and 1(b) with a proper normalization],
\begin{eqnarray}
G_E^{(Q)}(q^2) &=& Z_2 G_E^{(b)}(q^2) 
	\left ( \begin{array}{c} 1 \\ 0 \end{array} \right ) +
E_{NN}(q^2) \left ( \begin{array}{c}  {1/3} \\ {2/3}\end{array} \right ) +
E_{\Delta\Delta}(q^2) \left (\begin{array}{c}{4/3} \\ {-1/3}\end{array}\right) 
,\\
G_M^{(Q)}(q^2) &=& Z_2 G_M^{(b)}(q^2) 
	\left ( \begin{array}{c} 1 \\ -{2/3}\end{array} \right ) +
M_{NN}(q^2)\left ( \begin{array}{c}{1/27}\\ {-4/27}\end{array} \right )\nonumber\\
&+&
M_{\Delta\Delta}(q^2)\left ( \begin{array}{c}{20/27}\\ {-5/27}\end{array} 
\right ) +
M_{\Delta N}(q^2) \left ( \begin{array}{c} 
	{16\sqrt{2}/27} \\ {-16\sqrt{2}/27}\end{array} \right ),
\end{eqnarray}
where the upper and lower coefficients refer to 
the proton and neutron respectively. 
Here  $G_E^{(b)}$ and $G_M^{(b)}$ are the bare form factors calculated in 
Sec.~III and IV,
and  $E_{BC}$ and $M_{BC}$ are given by 
\begin{eqnarray}
E_{BC}(q^2) &=&  {f^{NB}f^{NC}\over 12\pi^2 m_\pi^2}
\int_0^\infty {dk\, k^4 u^2(k R)\over 
(\omega_{BN}+ \omega_k) (\omega_{CN}+\omega_k)\omega_k} \Omega_E(q,k),\\
M_{BC}(q^2) &=&  {f^{NB}f^{NC}\over 12\pi^2 m_\pi^2}
\int_0^\infty {dk\, k^4 u^2(k R)\over 
(\omega_{BN}+ \omega_k) (\omega_{CN}+\omega_k)\omega_k} \Omega_M(q,k),
\end{eqnarray}
where $\Omega_E(q,k)$ and $\Omega_M(q,k)$ contain the recoil corrections
for the intermediate baryons in Fig.~1(b) and are given by
\begin{eqnarray}
\Omega_E(q,k) &=& {1\over 4m_N}\int_{-1}^1\! dx 
\left[ (E'+m_N)(E+m_N)\right]^{1/2}
\left\{ 
\left [ 1 + {k^2 - q^2/4 \over (E'+m_N) (E+m_N)}\right] F_1(q^2)\right. 
\nonumber\\
&&\left.
-{1\over 2m_N}\left({q^2/2 +\vec{k}\cdot\vec{q} \over E'+m_N} 
        -{-q^2/2 +\vec{k}\cdot\vec{q} \over E+m_N}\right ) F_2(q^2)\right\},\\
\Omega_M(q,k) &=& {1\over 4m_N}\int_{-1}^1\! dx \left[ (E'+m_N)(E+m_N)
      \right]^{1/2}
\left\{ 
m_N\left ( {1\over E'+m_N} + {1\over E+m_N}\right ) F_1(q^2)\right. \nonumber\\
&&\left.
+ \left[ 1 + {k^2 - q^2/4 \over (E'+m_N) (E+m_N)}\right ] F_2(q^2)\right\}.
\end{eqnarray}
Without the recoil of the intermediate baryons 
(i.e., with $\vec{k}$ set to zero) 
the usual CBM results are thus recovered,
\begin{eqnarray}
\Omega_E(q,k) &=& F_1(q^2) - \eta F_2(q^2) = G_E^{(b)}(q^2), \\
\Omega_M(q,k) &=& F_1(q^2) + F_2(q^2) = G_M^{(b)}(q^2).
\end{eqnarray}

Now let us discuss the issue of  gauge invariance.
In the CBM the baryons are assumed to be  on-mass-shell,
thus it only makes sense to discuss current conservation as
a weak condition for electromagnetic gauge invariance.
In a static treatment, current conservation holds trivally.
With center-of-mass corrections, the electromagnetic form factors
are most conveniently calculated in the Breit frame. 
Since $q^0=0$ in this frame, current conservation is ensured
provided that 
\begin{equation}
\vec{q}\cdot
\bra{N({\vec{q}\over 2})}\vec{J}(0)\ket{N(-{\vec{q}\over 2})} =0. \label{weak}
\end{equation} 
For the quark core [Fig.~1(a)], explicit evaluations in both
PY and PT projection methods guarantee that the matrix element of the 
spatial component of the current is proportional to 
$\bfgreek{\sigma}\times\vec{q}$, and thus satisfies Eq.~(\ref{weak}).
For Fig.~1(b), as shown in the previous paragraph, 
all terms which are odd in $\vec{k}$ simply
vanish after the angular integration over the loop momentum, and the only
surviving term is proportional to $\bfgreek{\sigma}\times\vec{q}$,
therefore this diagram is separately gauge invariant. 
The proof of gauge invariance for Fig.~1(c) is slightly different since it is
evaluated in the heavy baryon approximation. 
By expanding the pion field in a plane wave and connecting the pion 
creation/annihilation operators with the CBM hamiltonian and the physical
baryon states, it is easy to show that\cite{TT83}
\begin{equation}
\vec{j}^{(\pi)}(0) \propto \int d\hat{k} 
\vec{k}\,\, \vec{k}\cdot\bfgreek{\sigma}\times\vec{q}= 
k^2\bfgreek{\sigma}\times\vec{q}.
\end{equation}
Thus this current is also transverse with respect to $\vec{q}$.
Since the total electromagnetic current is just the sum of the three 
contributions [Fig.~1(a), 1(b) and 1(c)] in the CBM, and hence current 
conservation, Eq.~(\ref{weak}), is satisfied in this calculation.
\end{appendix}

\begin{table}
\caption{Magnetic moments of the nucleon. The static case  refers to 
the original CBM results without center of mass correction, and PY and PT 
are for two calculations with momentum projected wave functions.
The experimental values are $2.79\mu_B$ and $-1.91\mu_B$, respectively.  
}
\label{mag1.02}
\begin{center}
\begin{tabular}{l|ccc|ccr}
 & \multicolumn{3}{c|}{proton}   &  \multicolumn{3}{c}{neutron}  \\ \hline
 R(fm) & static & PY & PT &  static & PY & PT  \\ \hline
 0.8 & 2.49 & 2.25 & 2.36 & -2.06 & -1.89 & -1.97 \\
 0.9 & 2.44 & 2.18 & 2.30 & -1.96 & -1.78 & -1.86 \\  
 1.0 & 2.46 & 2.18 & 2.31 & -1.92 & -1.73 & -1.81 \\ 
 1.1 & 2.53 & 2.23 & 2.36 & -1.93 & -1.71 & -1.81 \\
\end{tabular}
\end{center}
\end{table}

\begin{figure}
\vspace{2.0cm}
\centering{\
\epsfig{file=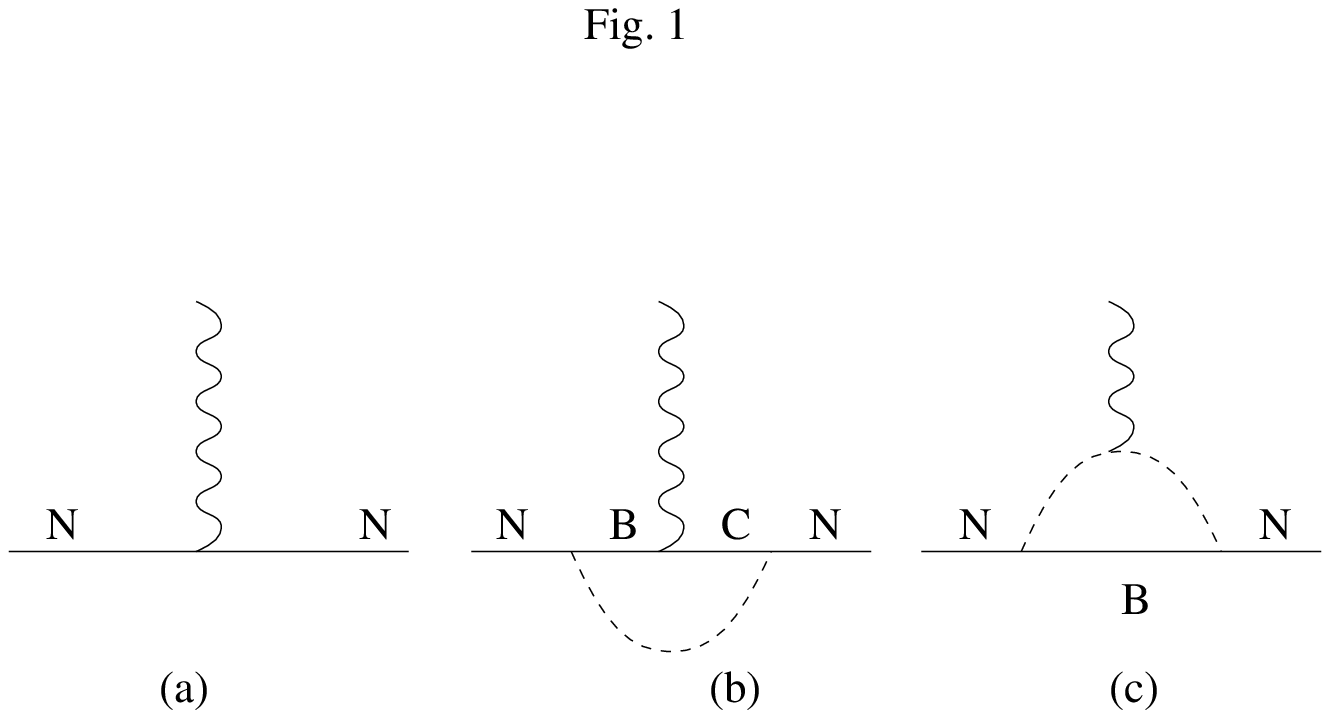}
\vspace*{1.0cm}
\caption{Diagrams illustrating the various contributions included in
this calculation (up to one pion loop). The intermediate baryons $B$ 
and $C$ are restricted to the $N$ and $\Delta$.}}
\label{fig.feyn}
\end{figure}

\newpage
\vspace*{1cm}
\begin{figure}
\centering{\
\epsfig{file=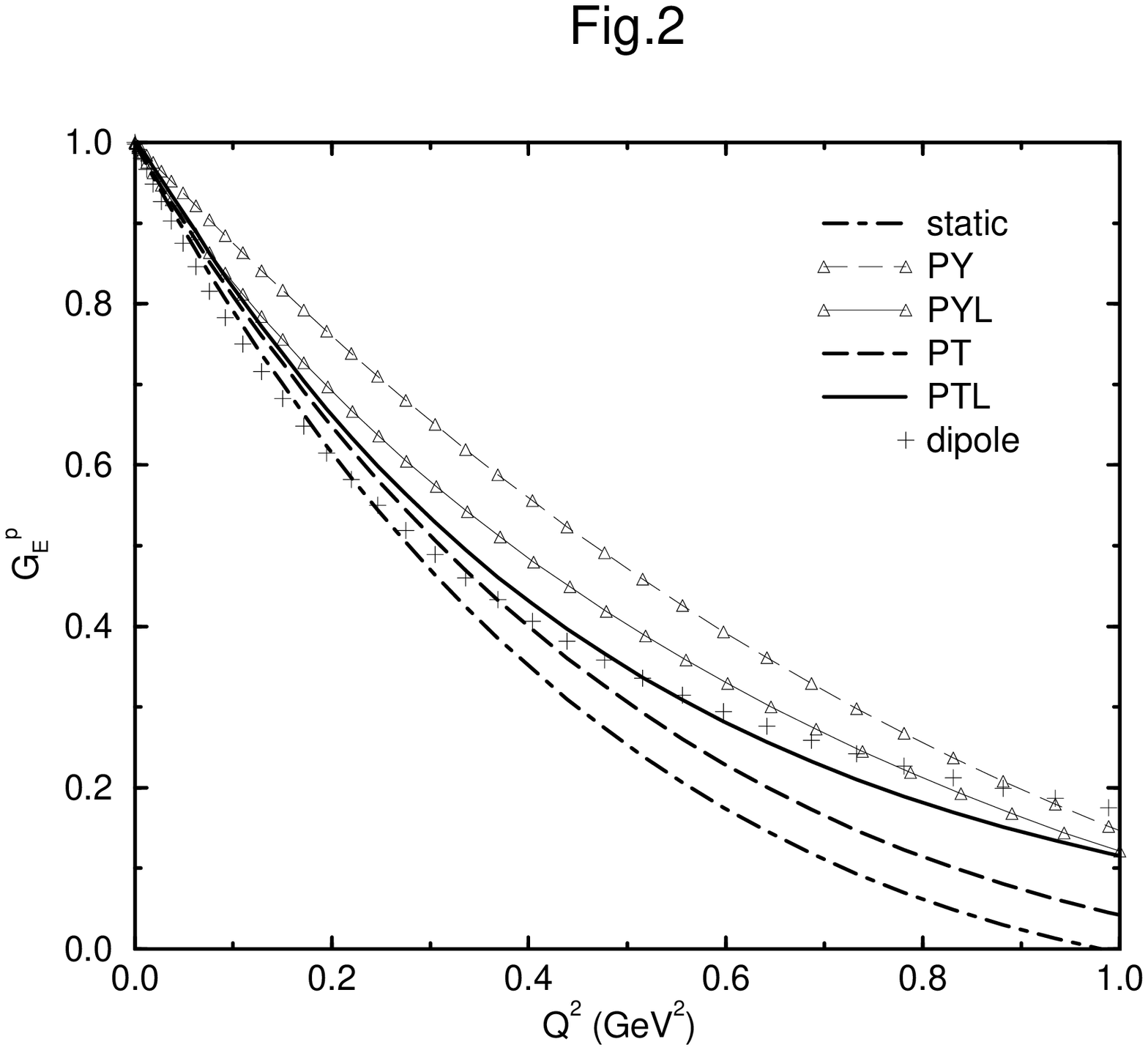,height=10cm,width=12cm}
\caption{The effect of the center of mass correction
and Lorentz contraction for the charge form factor of the bare proton. 
The bag radius is taken to be 1.0 fm. The ``static'' curve refers to the 
naive MIT cavity approximation, PY and PT stand for Peierls--Yoccoz 
and Peierls-Thouless projection, respectively, and PYL and PTL 
for the corresponding versions with the Lorentz contraction.} 
\label{fig2.ps}}
\end{figure}

\newpage
\vspace*{1cm}
\begin{figure}
\centering{\
\epsfig{file=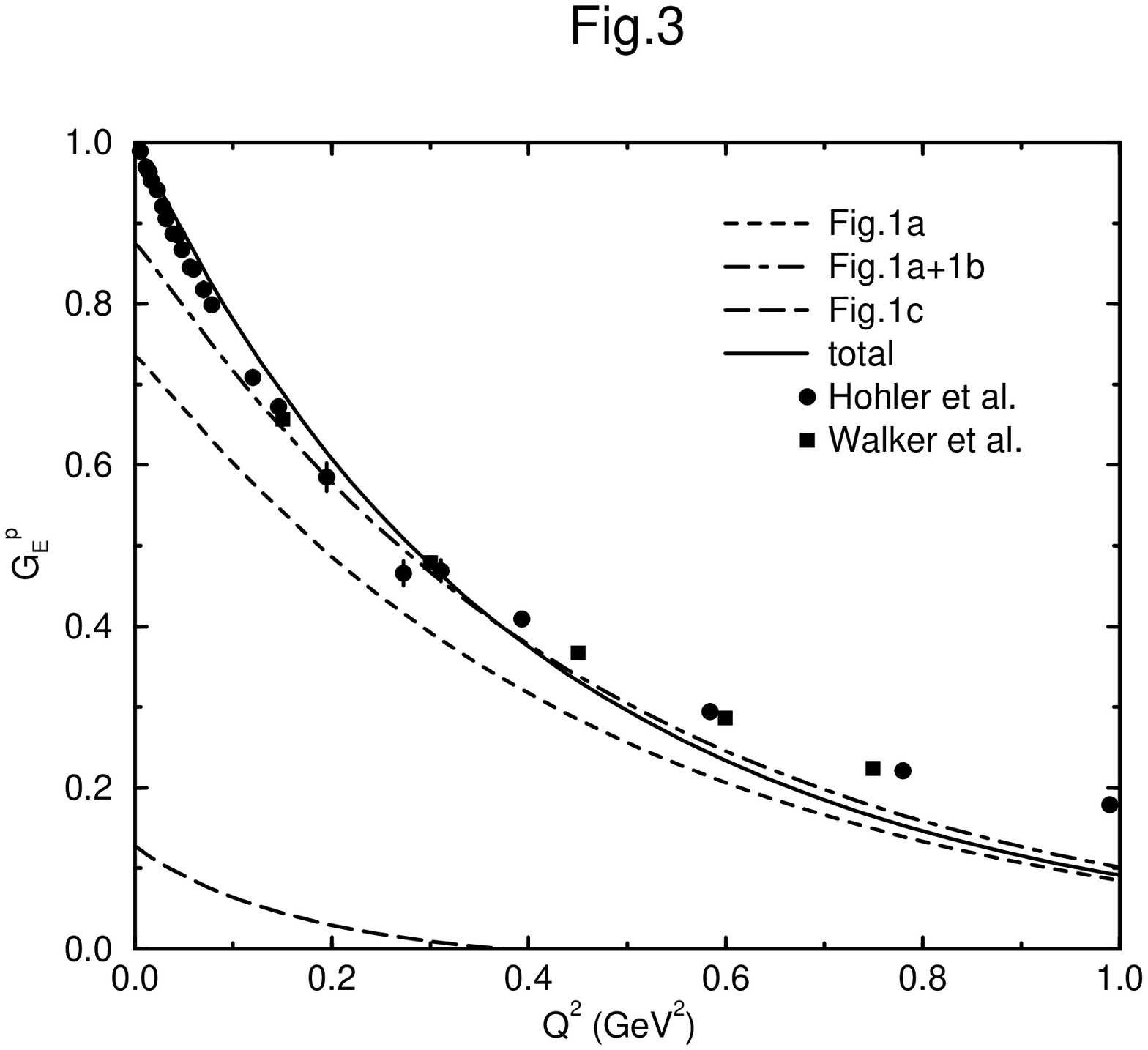,height=10cm,width=12cm}
\caption{The individual contributions to the proton  charge form factor
with the bag radius R = 1.0 fm. The quark part is calculated using
the Lorentz contracted PT wave functions. 
Experimental data are taken from Ref.~\protect{\cite{ep}}.}
\label{fig3.ps}}
\end{figure}

\newpage
\vspace*{1cm}
\begin{figure}
\centering{\
\epsfig{file=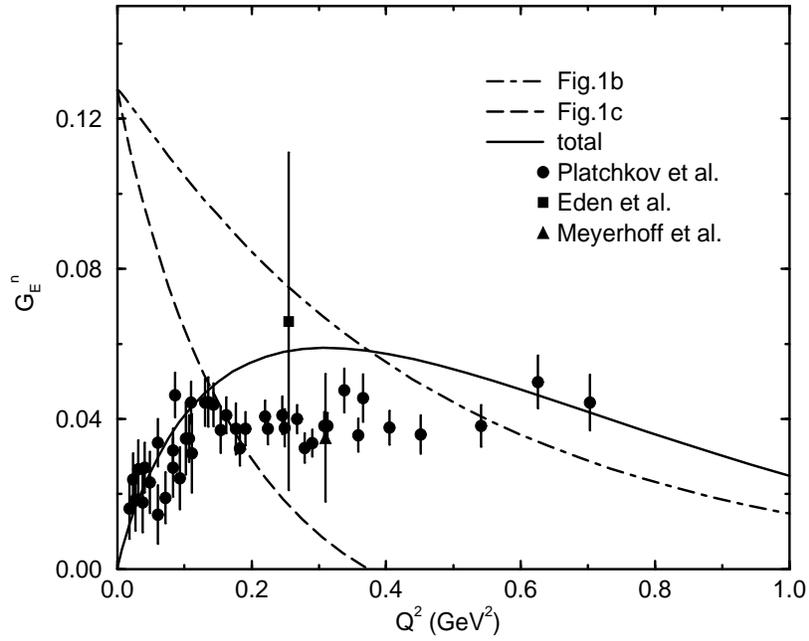,height=10cm,width=12cm}
\caption{The individual contributions to the  neutron charge form factor. 
The key is as in Fig.~3, except that the experimental data is from 
Ref.~\protect{\cite{en}}. As the contribution from Fig.~1(c) is negative, 
we show its magnitude for convenience.}
\label{fig4.ps}}
\end{figure}

\newpage
\vspace*{1cm}
\begin{figure}
\centering{\
\epsfig{file=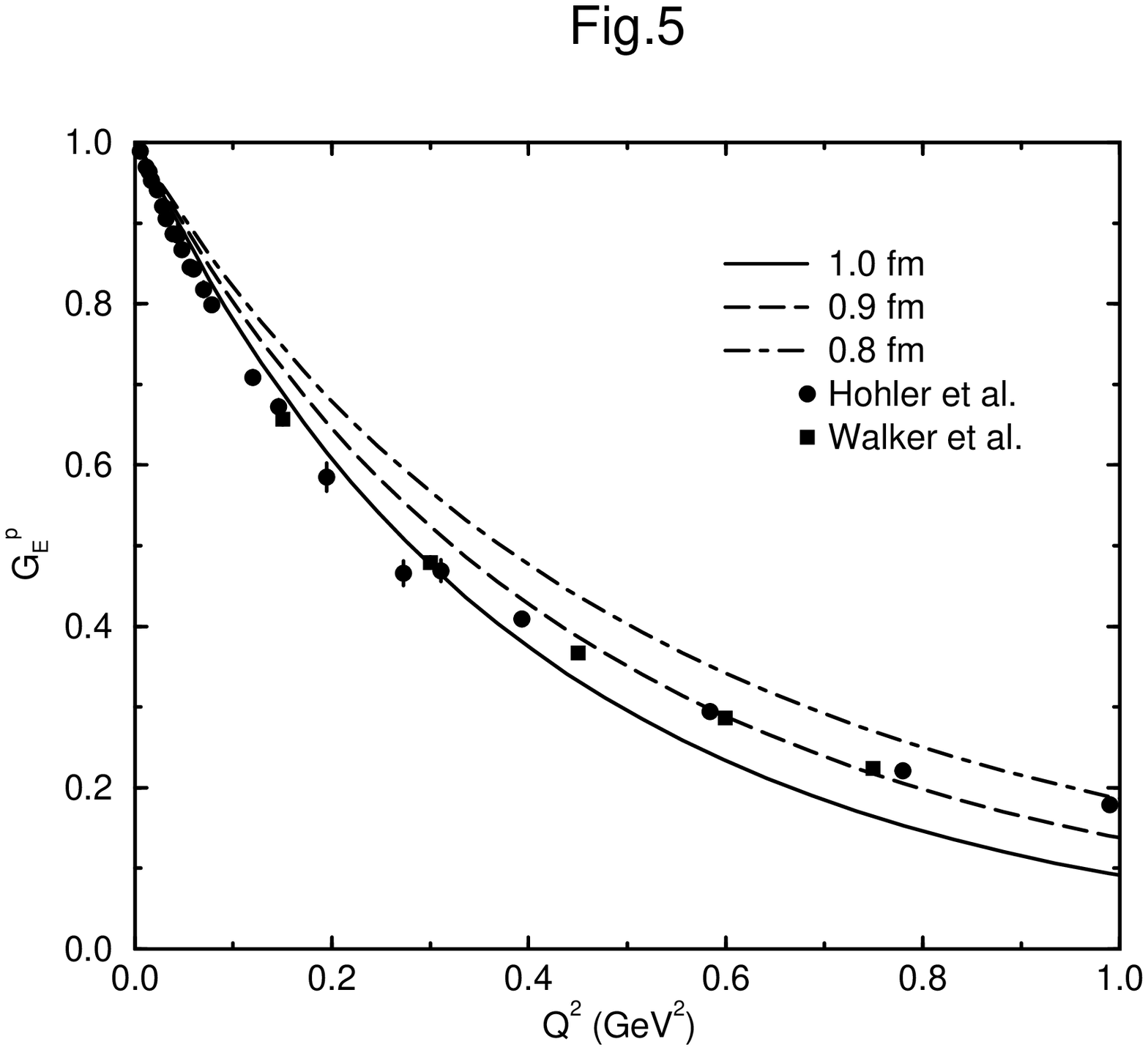,height=10cm,width=12cm}
\caption{The proton charge form factor for three different bag radii. 
Lorentz contracted PT wave functions (with $\eta_p=1$) 
are  used in the calculations. 
Data are the same as in Fig.~3.}
\label{fig5.ps}}
\end{figure}

\newpage
\vspace*{1cm}
\begin{figure}
\centering{\
\epsfig{file=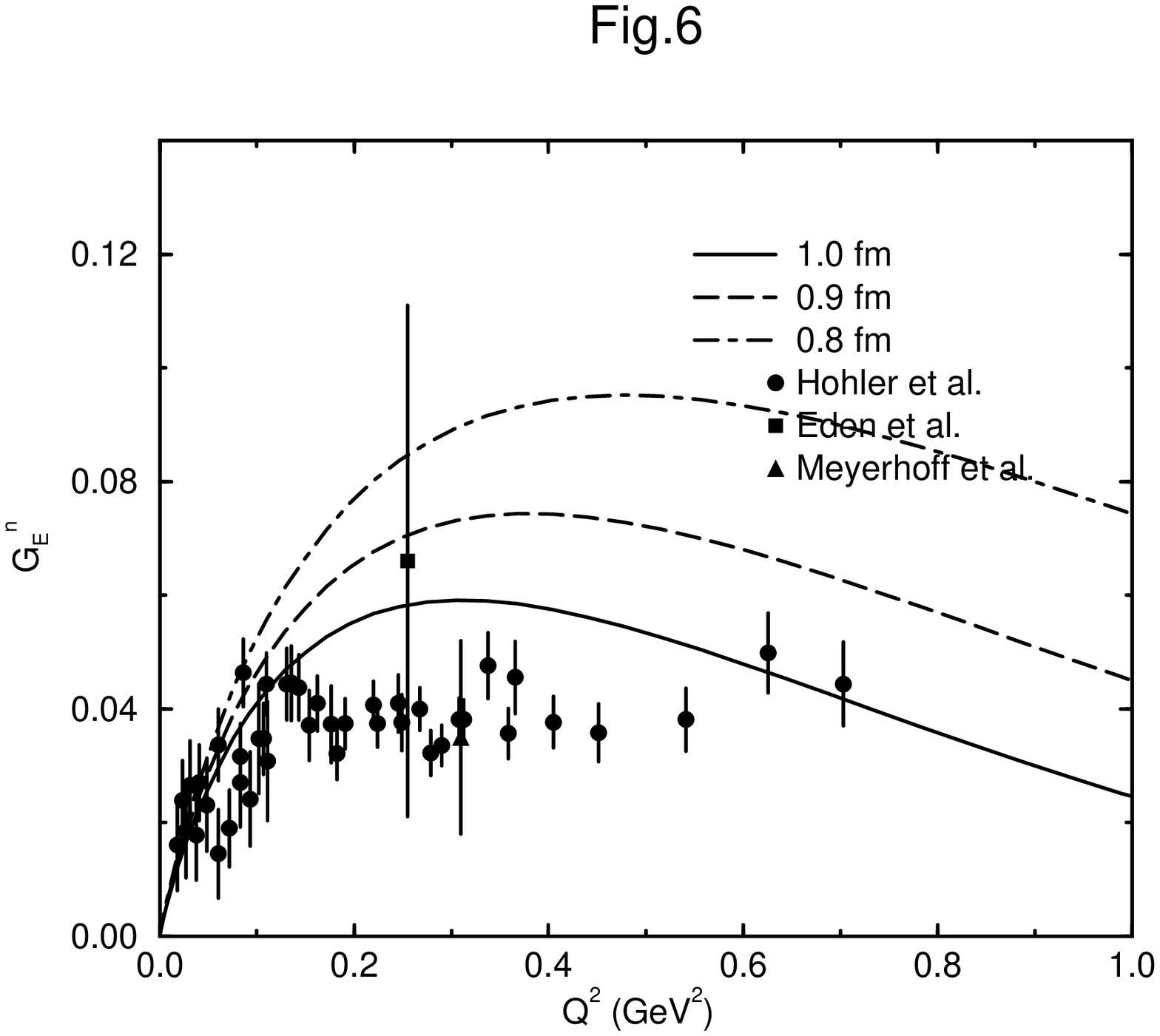,height=10cm,width=12cm}
\caption{The neutron charge form factor 
using Lorentz contracted PT wave functions. 
Data are the same as in Fig.~4. }
\label{fig6.ps}}
\end{figure}

\newpage
\vspace*{1cm}
\begin{figure}
\centering{\
\epsfig{file=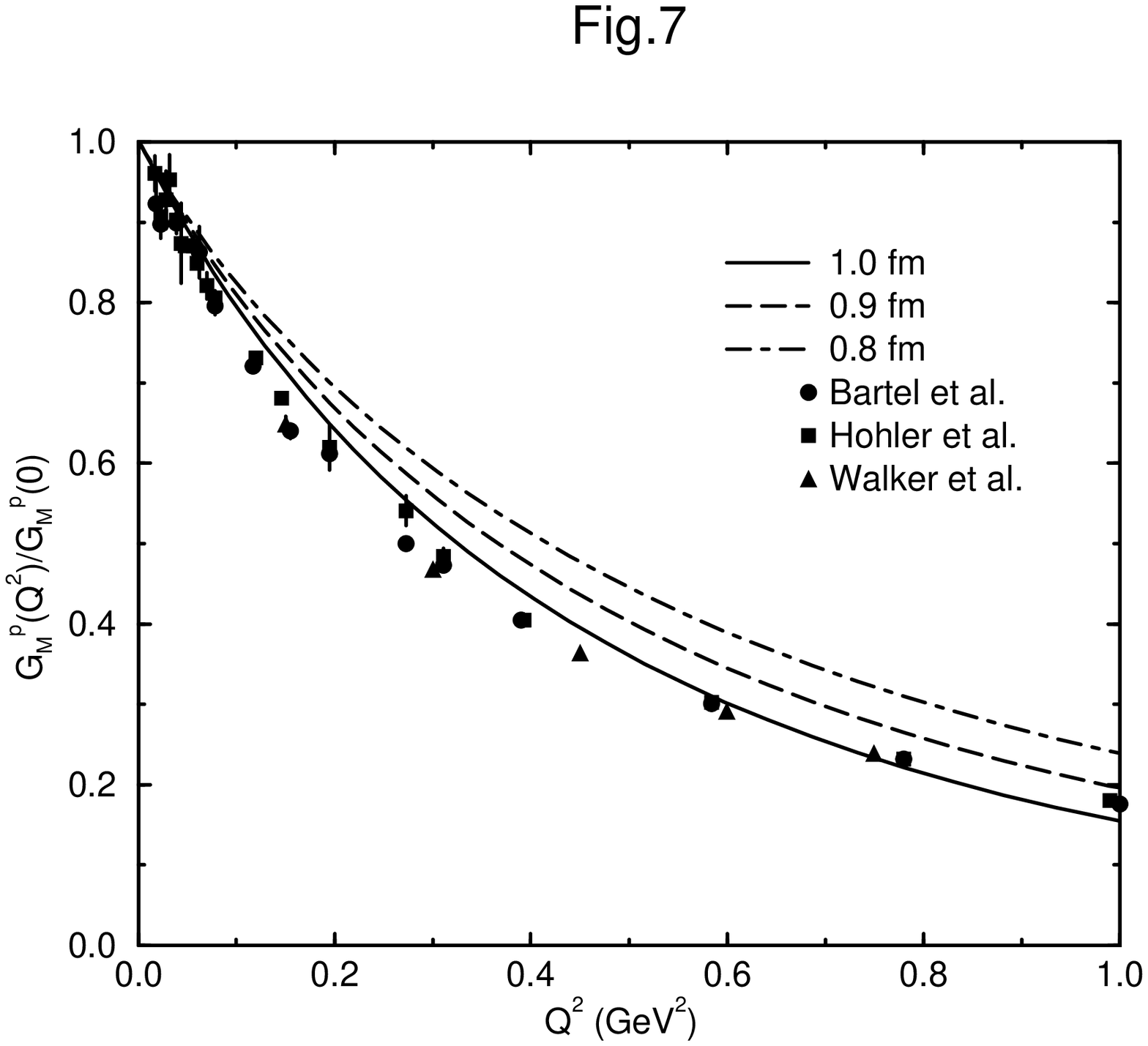,height=10cm,width=12cm}
\caption{The proton magnetic form factor 
using Lorentz contracted PT wave functions. 
Experimental data are from Ref.~\protect{\cite{ep,mp}}. }
\label{fig7.ps}}
\end{figure}

\newpage
\vspace*{1cm}
\begin{figure}
\centering{\
\epsfig{file=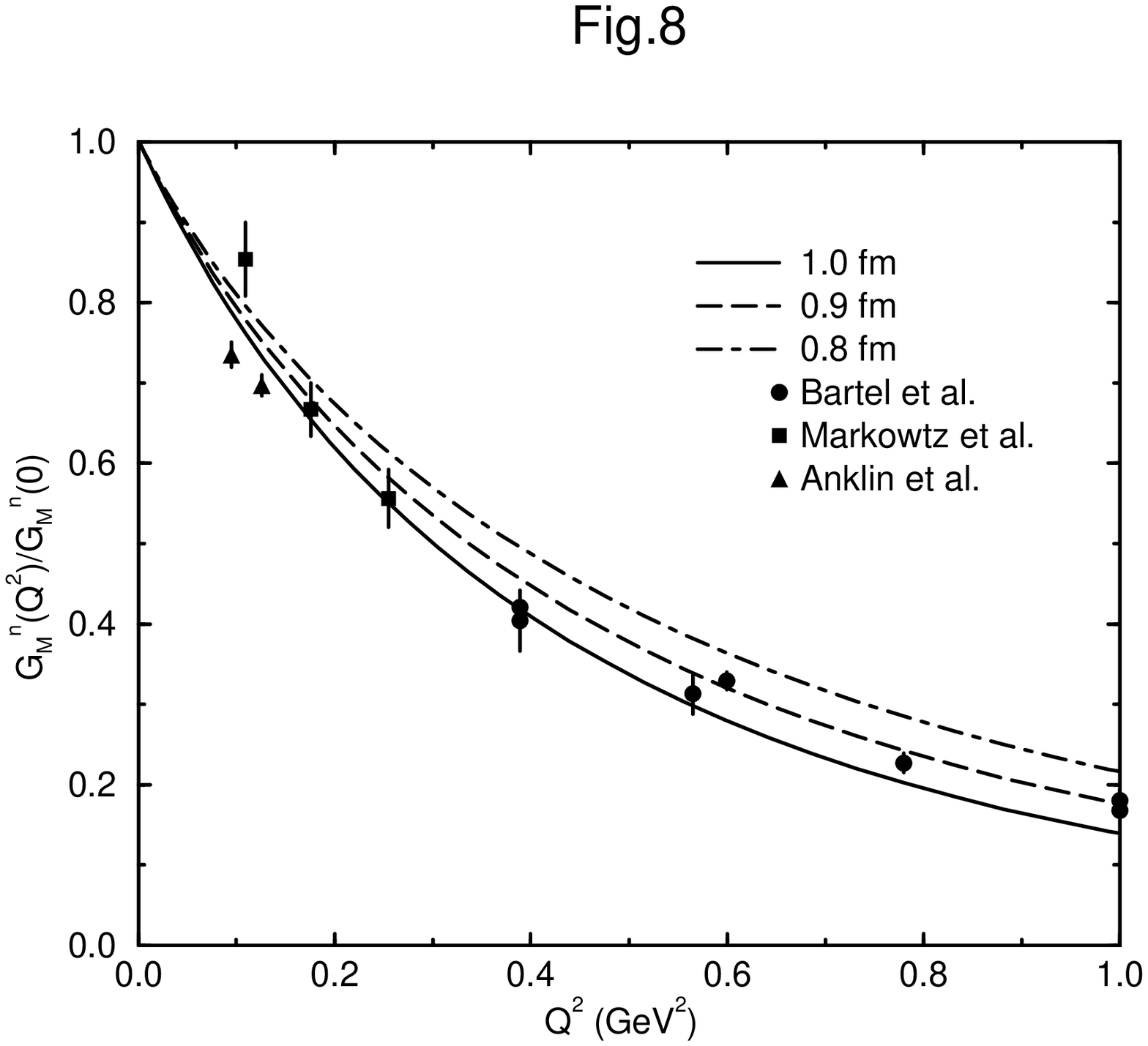,height=10cm,width=12cm}
\caption{The neutron magnetic form factor
using Lorentz contracted PT wave functions. 
Experimental data are from Ref.~\protect{\cite{mp,mn}}. }
\label{fig8.ps}}
\end{figure}


\begin{references}
\bibitem{MIT} A. Chodos, R. L. Jaffe, K. Johnson, C. B. Thorn 
        and V. F. Weisskopf, Phys. Rev. D {\bf 9}, 3471 (1974);
              A. Chodos, R. L. Jaffe, K. Johnson and C. B. Thorn,
                {\em ibid.} {\bf 10}, 2599 (1974);
              T. A. DeGrand, R. L. Jaffe, K. Johnson and J. Kiskis,
                {\em ibid.} {\bf 12}, 2060 (1975).
\bibitem{CBM} A. W. Thomas, Adv. Nucl. Phys. {\bf 13}, 1 (1984);
              G. A. Miller, Int. Rev. Nucl. Phys., {\bf 2}, 190 (1984).
\bibitem{medium97} D. H. Lu, A. W. Thomas, K. Tsushima, A. G. Williams,
        and K. Saito, nucl-th/9706043, to appear in Phys. Lett. B.
\bibitem{WILETS}
        L. Wilets, {\em Non-Topological Solitons}
        (World Scientific, Singapore, 1989); 
        M. C. Birse, Prog. Part. Nucl. Phys. {\bf 25}, 1 (1990).
\bibitem{BETZ83}
	M. Betz and R. Goldflam, Phys. Rev. D {\bf 28}, 2848 (1983).
\bibitem{ANTWERPEN94}
	C. H. M. Antwerpen, Ph.~D. thesis (Flinders University, 1994). 
\bibitem{DJ80}J. Donoghue and K. Johnson, Phys. Rev. D {\bf 21}, 1975 (1980).
\bibitem{WONG81}C. W. Wong, Phys. Rev. D {\bf 24}, 1416 (1981).
\bibitem{TEGEN82} R. Tegen, R. Brockmann, and W. Weise, 
	Z. Phys. A {\bf 307}, 339 (1982); 
	E. Oset, R. Tegen, and W. Weise, Nucl. Phys. {\bf A426}, 456 (1984).
\bibitem{Tegen90} R. Tegen, 
	Ann. Phys. (N.Y.) {\bf 197}, 439 (1990), and references therein.
\bibitem{PY57}
        R. E. Peierls and J. Yoccoz, Proc. Phys. Soc. London
        {\bf A70},  381 (1957).
\bibitem{PT62}
        R. E. Peierls and D. J. Thouless, Nucl. Phys. {\bf 38}, 154 (1962).
\bibitem{DELTA} 
	D. H. Lu, A. W. Thomas, and A. G. Williams, 
	Phys. Rev. C {\bf 55}, 3108 (1997).
\bibitem{LP70}
	A. L. Licht and A. Pagnamenta, Phys. Rev. D {\bf 2}, 1156 (1970).
\bibitem{TT83} S. Th\'eberge, G. A. Miller, and A. W. Thomas, Can. J. Phys.
                {\bf 60}, 59 (1982); 
	S. Th\'eberge and A. W. Thomas, 
	Nucl. Phys. {\bf A393}, 252 (1983); 
	S. Th\'eberge, Ph.~D. thesis 
	(University of British Columbia, 1982).
\bibitem{LC}see e.g., P. L. Chung and F. Coester, 
	Phys. Rev. D {\bf 44}, 229 (1991); 
	S. Capstick and B. Keister, Phys. Rev. D {\bf 51}, 3598 (1995);
	F. Cardarelli, E. Pace, G. Salme, and S. Simula,
	Nucl. Phys. {\bf A623}, 361 (1997), and references therein.
\bibitem{Jaffe81} R. L. Jaffe, Ann. Phys. (N.Y.) {\bf 132}, 32 (1981). 
\bibitem{JI91} X. Ji, Phys. Lett. B {\bf 254} (1991); 
	G. Holzwarth, Z. Phys. A {\bf 356}, 339 (1996). 
\bibitem{DODD81} L. R. Dodd, A. W. Thomas and R. F. Alvarez-Estrada, 
	Phys. Rev. D {\bf 24}, 1961 (1981).
\bibitem{Gross} F. Gross and D. O. Riska, 
	Phys. Rev. C {\bf 36}, 1928 (1987);
	K. Ohta, Phys. Rev. C {\bf 40}, 1335 (1989);
	W. Koepf and E.M. Henley, 
	Phys. Rev. C {\bf 49}, 2219 (1994).
\bibitem{Miller97} G. A. Miller and A. W. Thomas,
	Phys. Rev. C {\bf 56}, 2329 (1997).
\bibitem{KAZUO89}T. Yamaguchi, K. Tsushima, Y. Kohyama and K. Kubodera,
	Nucl. Phys. {\bf A500}, 429 (1989).
\bibitem{Morgan86} M. A. Morgan, G. A. Miller, and A. W. Thomas,
	Phys. Rev. D {\bf 33}, 817 (1986). 
\bibitem{nrms} A. Berard {\em et al.}, Phys. Lett. B {\bf 47}, 355 (1973);
	V. E. Krohn and R. Ringo, Phys. Rev. D {\bf 8}, 1305 (1973);
	L. Koester {\em et al.}, Phys. Rev. Lett. {\bf 36}, 1021 (1976);
	G. G. Bunatian {\em et al.}, Z. Phys. {\bf A359}, 337 (1997).
\bibitem{ep} 
	G. Hohler {\em et al.}, Nucl. Phys. {\bf B114}, 505 (1976);
	R. C. Walker {\em et al.}, Phys. Rev. C {\bf 49}, 5671 (1994).
\bibitem{en} 
	S. Platchkov {\em et al.}, Nucl. Phys. {\bf A510}, 740 (1990);
	T. Eden {\em et al.}, Phys. Rev. C {\bf 50}, 1749 (1994);
	M. Meyerhoff {\em et al.}, Phys. Lett. B {\bf 327}, 201 (1994).
\bibitem{mp} W. Bartel {\em et al.}, Nucl. Phys. {\bf B58}, 429 (1973).
\bibitem{mn}
	P. Markowtz {\em et al.}, Phys. Rev. C {\bf 48}, 5 (1993);
	H. Anklin {\em et al.}, Phys. Lett. B {\bf 336}, 313 (1994).
\end{references}
\end{document}